\documentclass[a4paper,english,11pt]{article}
\usepackage[dvips]{graphicx,epsfig,color}
\usepackage{wrapfig,rotating}
\usepackage{amssymb,amsmath,array}
\usepackage{subfigure}
\usepackage{wrapfig}
\usepackage{sidecap}
\usepackage{geometry}

\begin{document}
\title{Predictions for proton-proton interaction cross-sections at LHC}

\author{{\slshape \fbox{A.B. Kaidalov}$^{1}$, M.G. Poghosyan$^{2}$}\\[1ex]
$^{1}$ Institute of Theoretical and Experimental Physics, 117526 Moscow, Russia\\
$^{2}$ Universit\`a di Torino and INFN, 10125 Torino, Italy
}
\date{}
\maketitle

\begin{abstract}
This is a short communication with a summary of results obtained with a model 
based on Gribov's reggeon calculus, which was proposed and applied to processes 
of soft diffraction at high energies. We present a brief description of the model 
and its predictions for various LHC energies. 
\end{abstract}

\begin{figure}[h!]
  \centering
\includegraphics[width=0.7\textwidth]{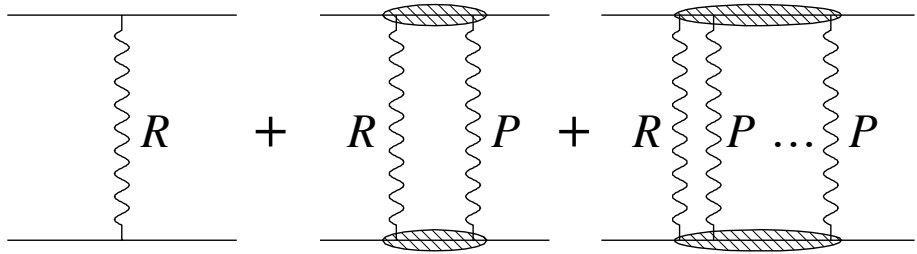}
  \caption{
Single Regge-pole and multi-Pomeron exchange diagrams. The inclusion of 
the multi-Pomeron exchange (eikonalisation) suppresses the fast growth 
of the total cross-section which is expected from Pomeron pole exchange, 
and restores unitarity.}
\label{Fig:MultiReggeExchange}
\end{figure}

\begin{figure}[h!]
  \centering
\includegraphics[width=0.4\textwidth]{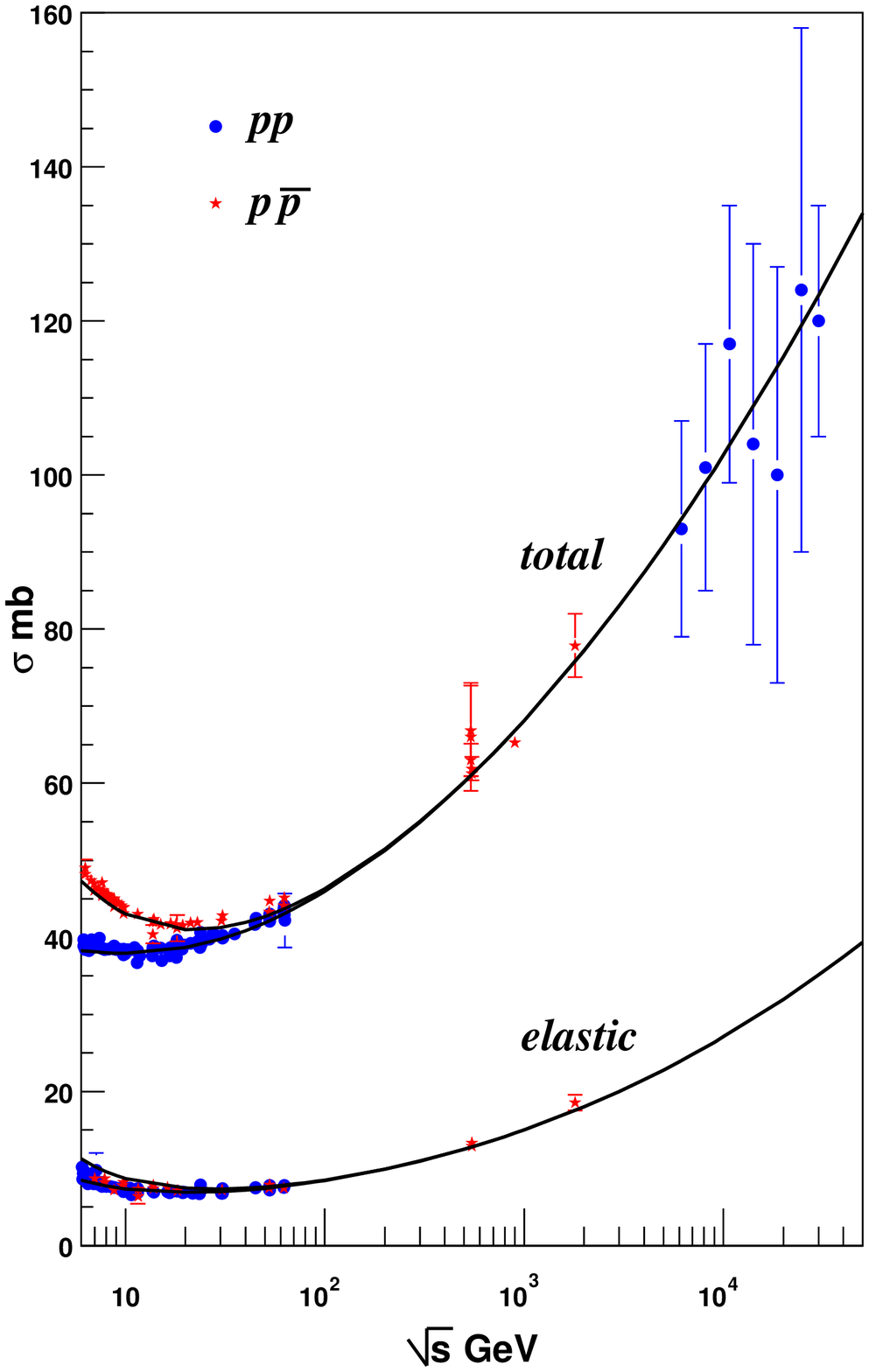}
\includegraphics[width=0.4\textwidth]{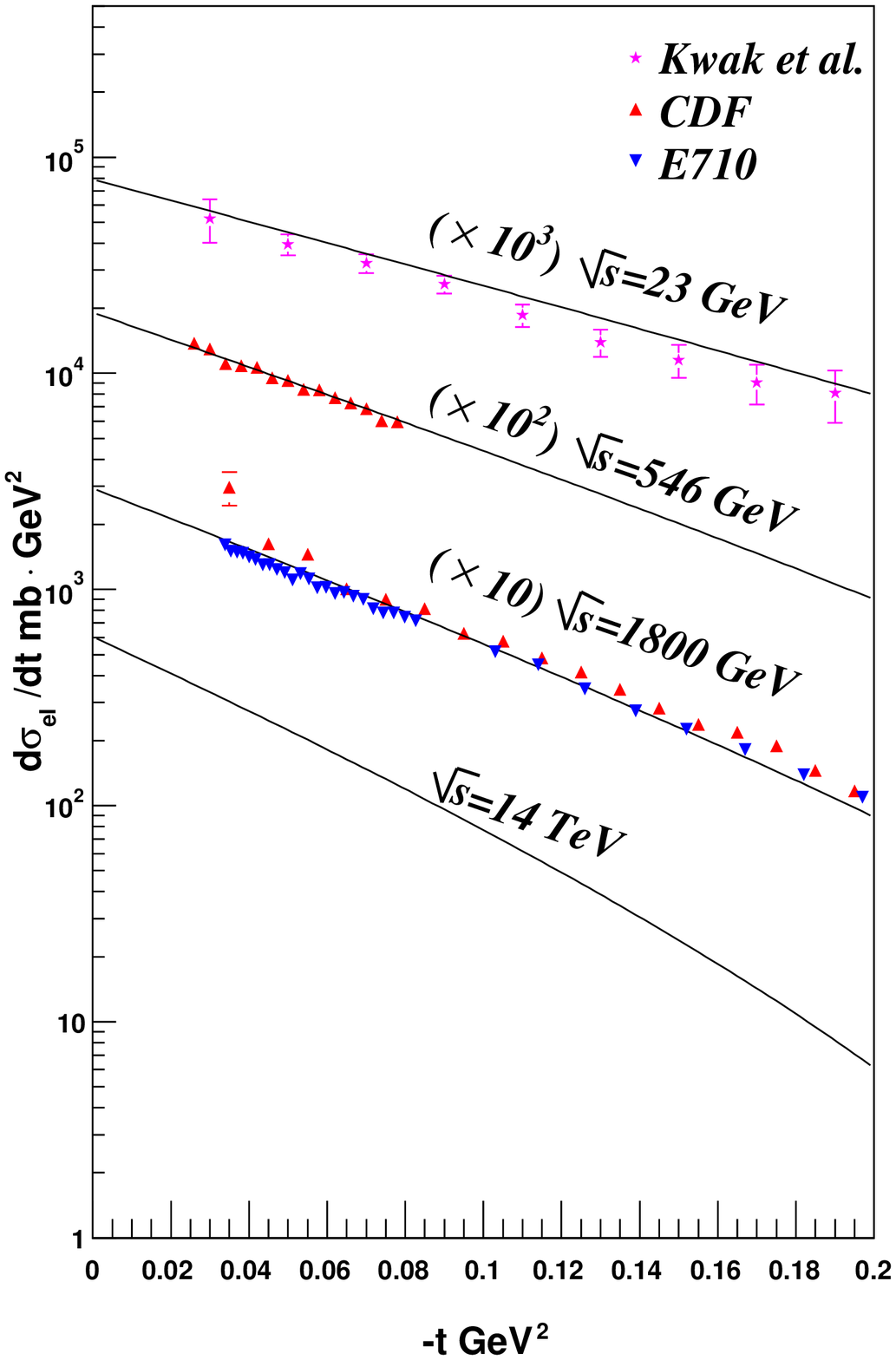}
  \caption{
Comparison of the fit result with data on $pp$ and $p\bar{p}$ total and elastic cross-section \cite{XS}.
}
\label{Fig:TotalElastic}
\end{figure}

\begin{figure}[h!]
  \centering
\includegraphics[width=0.37\textwidth]{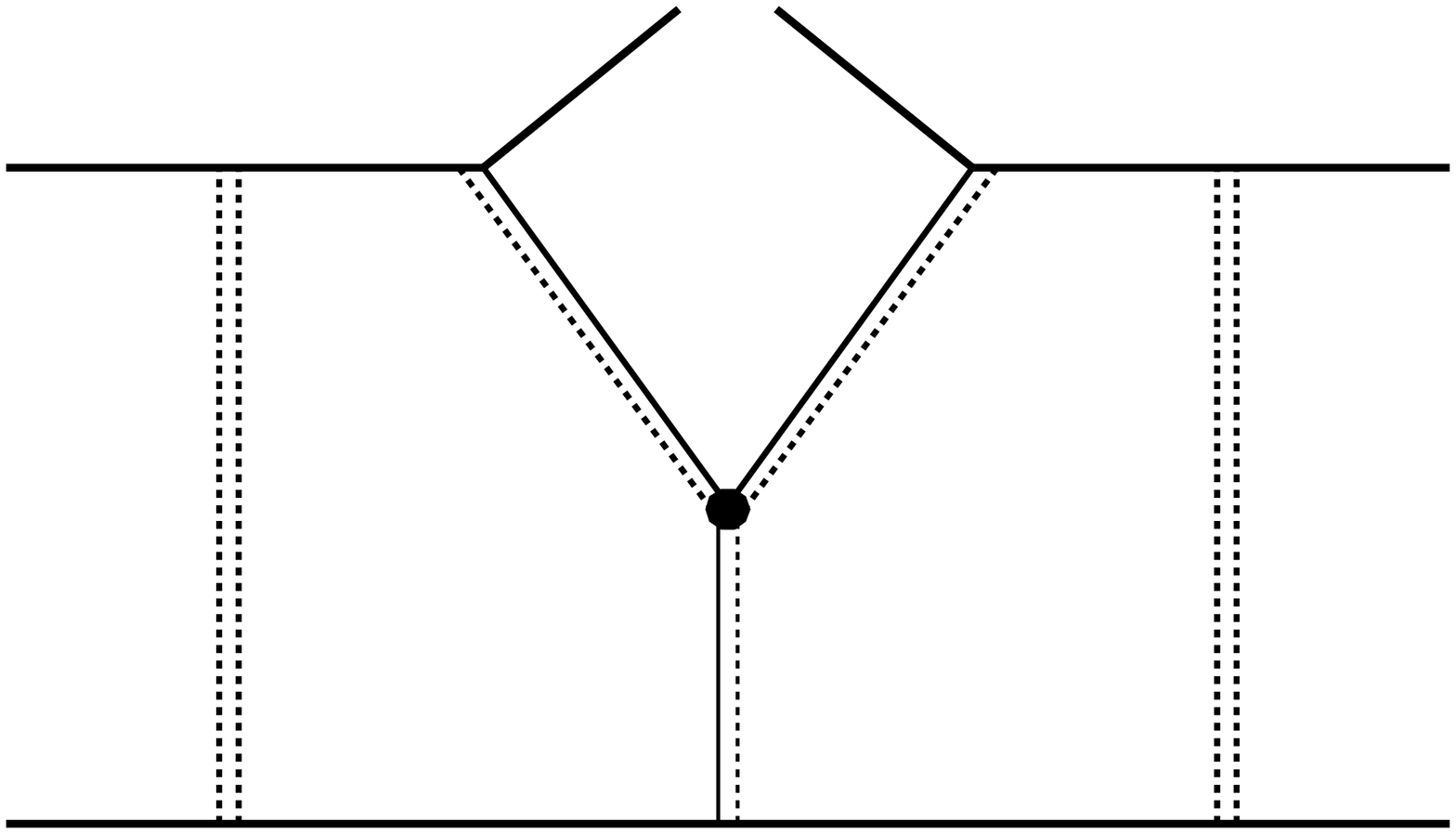}
\includegraphics[width=0.37\textwidth]{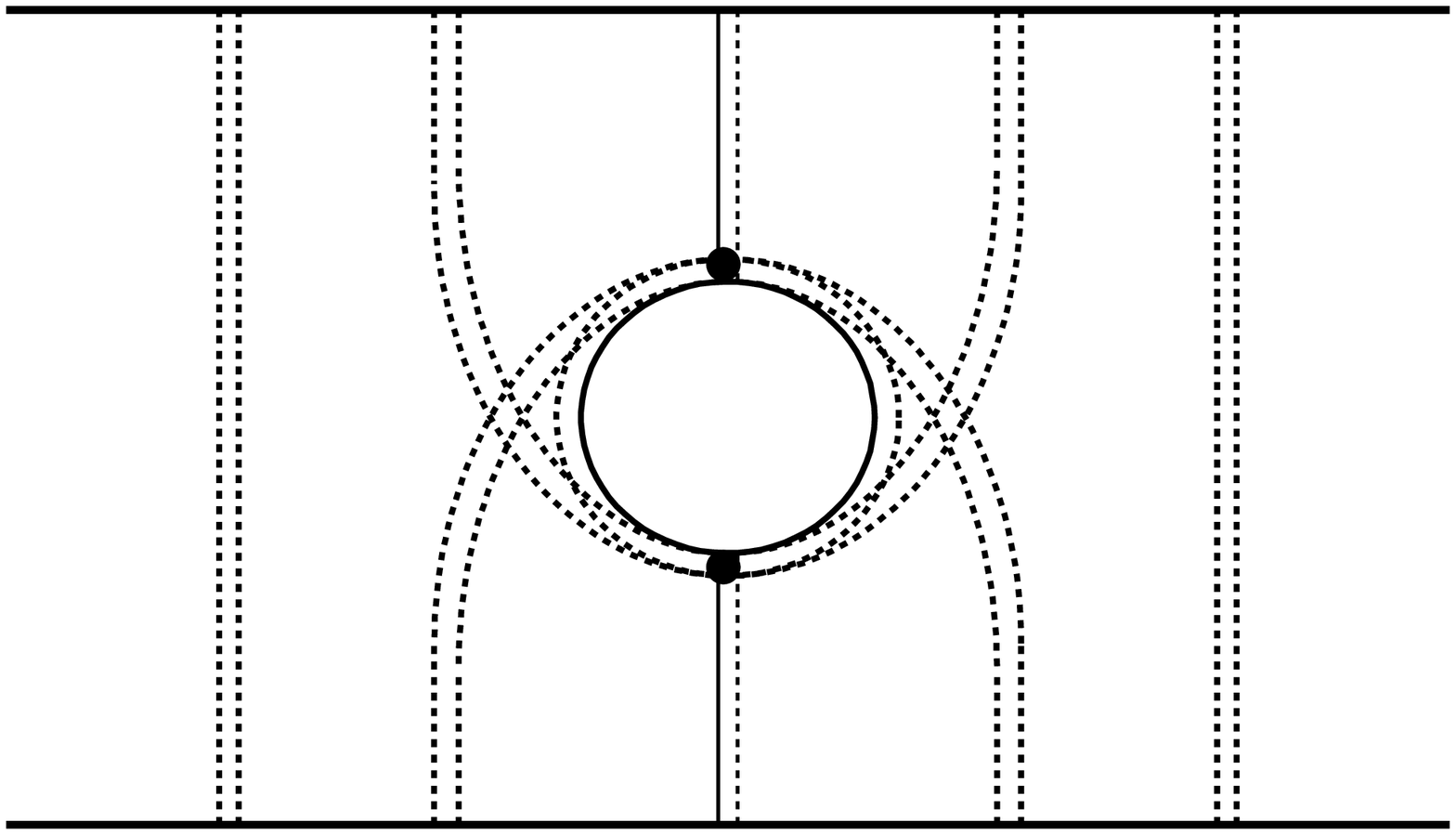}
  \caption{
Eikonalised triple-Reggeon and loop diagrams used for describing the single-diffraction 
dissociation process in hadronic collisions. The solid lines correspond to Pomeron and Reggeon 
exchanges and the dotted lines correspond to any number Pomeron exchange.
}
\label{Fig:SD_DDdiag}
\end{figure}

\begin{figure}[h!]
  \centering
\includegraphics[width=0.45\textwidth]{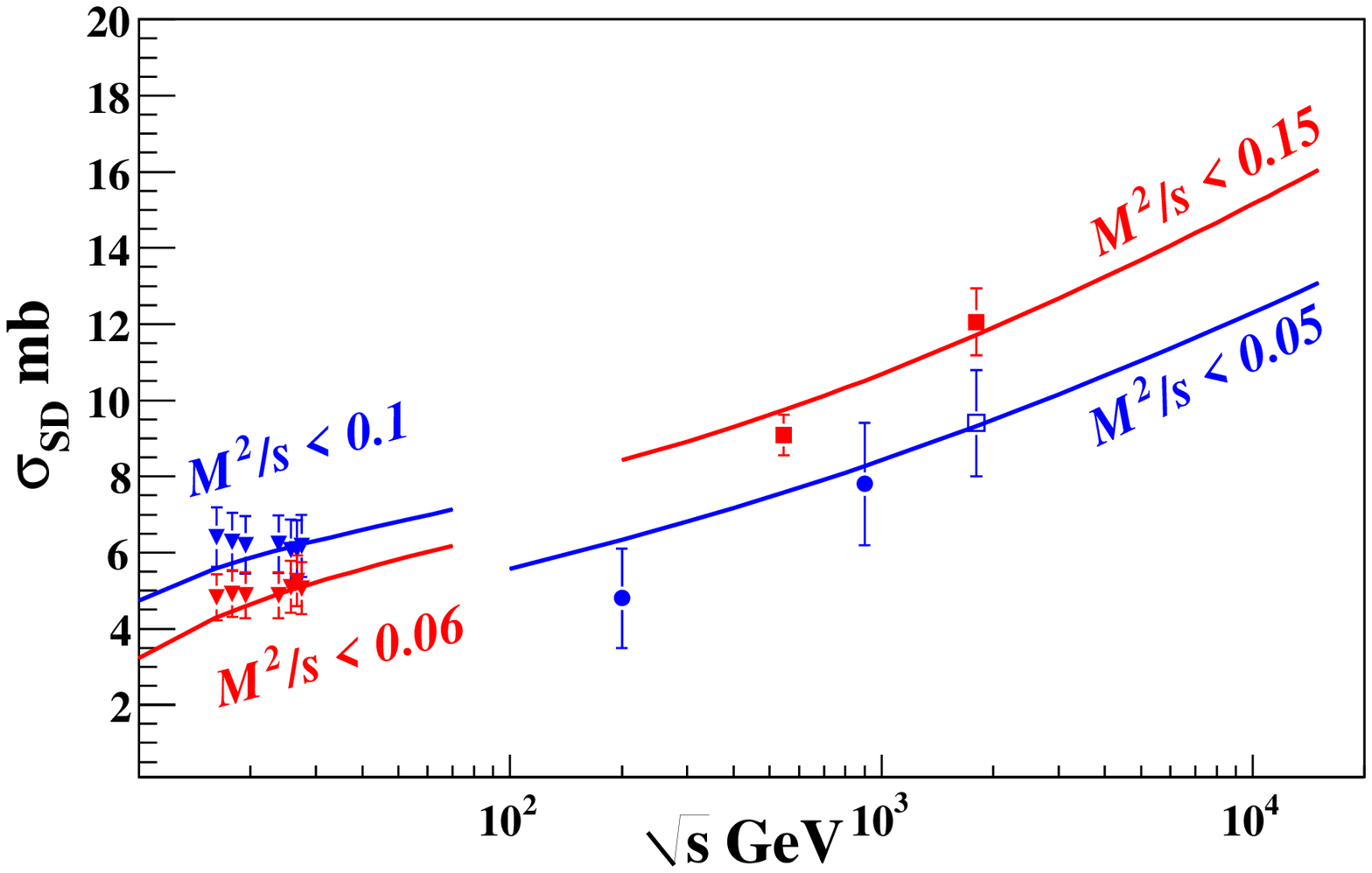}
\includegraphics[width=0.45\textwidth]{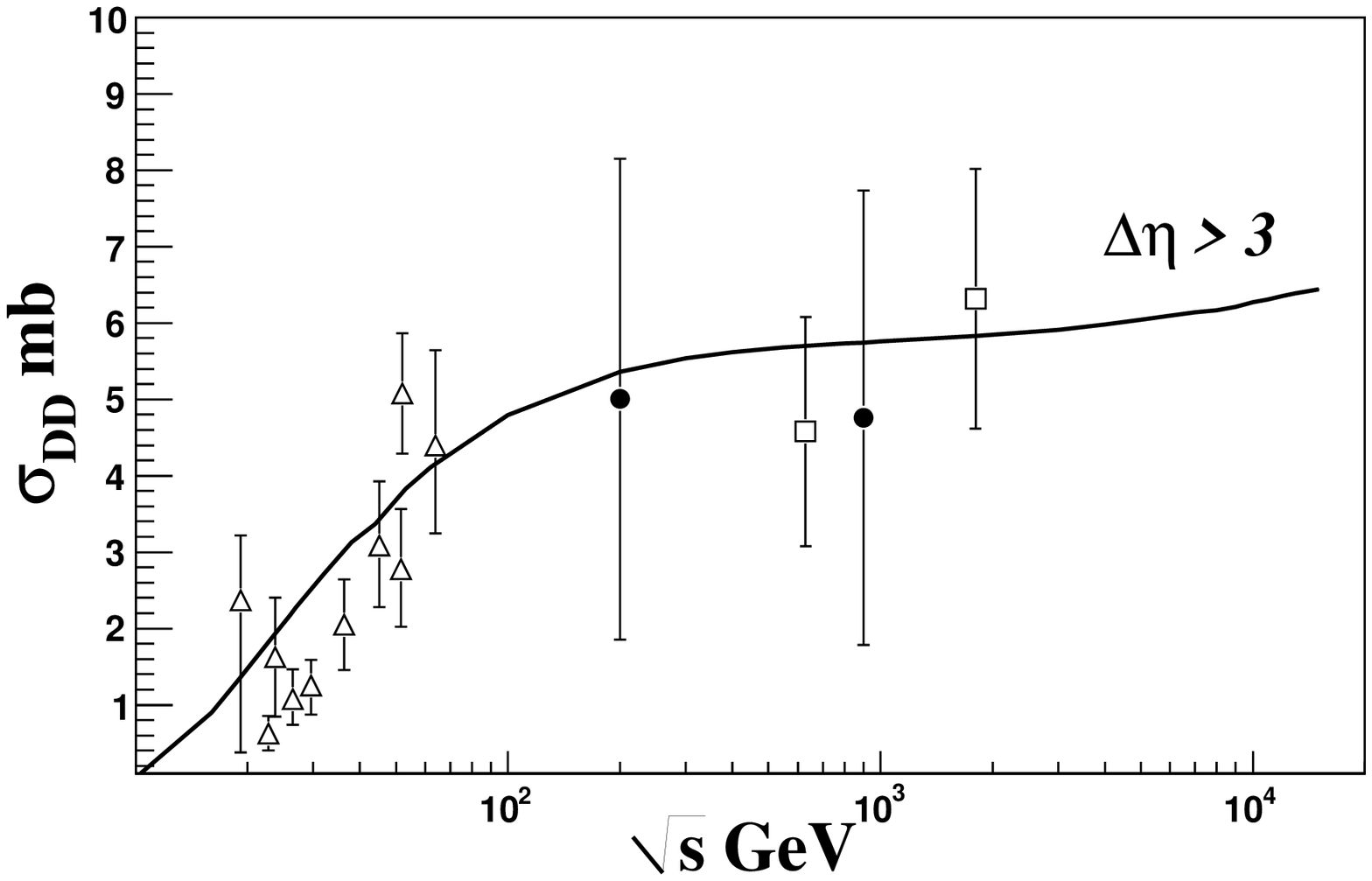}
  \caption{
Integrated single- (right) and double- (left) diffractive cross-section as a function of $\sqrt{s}$. 
The integrations are done in accordance with corresponding measurement as they are indicated in the plots. 
Data are taken from \cite{SD_DD}.
}
\label{Fig:SD_DD}
\end{figure}

Regge theory \cite{Coll} is the main method for describing high-energy soft processes. 
The asymptotic behavior of the cross-sections of elastic scattering and multiple productions 
of hadrons is determined by the properties of the Pomeron, the right-most pole singularity of 
the elastic amplitude in the complex-momentum plane ($j$-plane). The experimentally observed 
increase of the total cross-section with increasing energy makes it necessary to consider the 
Pomeron with intercept $\alpha_P = 1+ \Delta > 1$ ($\sigma_{tot} \sim s^{\Delta}$). The inclusion 
of only the Pomeron pole contribution in the elastic scattering amplitude leads to violation 
of the unitarity. This difficulty is removed by taking into account the contribution of branch 
points of the amplitude in the $j$-plane, which corresponds to the multi-Reggeon exchange in the 
$t$-channel. A Regge-pole exchange can be interpreted as a single scattering while branch points 
correspond to multiple scatterings on the constituents of hadrons.\\
The contribution of the branch points involved in the exchange of several Reggeons associated with 
$\rho, f, \omega,$ etc. decreases very quickly with increasing collision energy and the contribution 
of such branch points can be neglected with respect to the branch points due to the exchange of one 
of them and any number of Pomerons. In the eikonal approximation, where the interaction between 
Pomerons is neglected, the elastic scattering amplitude can be parametrized by the sum of diagrams shown in
Fig.~\ref{Fig:MultiReggeExchange}.\\
In these graphs, the wavy lines labeled with $R$ correspond to Pomeron and Reggeon exchanges, 
whereas the wavy lines labeled with $P$ correspond only to Pomeron exchange. The account of 
the multi-Pomeron exchange restores unitarity of the 
scattering amplitude, which leads to a behavior of the total cross-section for $s >> m_{N}^{2}$ : 
$\sigma_{tot} \sim \ln ^2 s$, which satisfies the Froissart bound.\\
For the analysis of $pp$ and $p\bar{p}$ total and elastic cross-section  data we saturate 
the matrix elements by the contribution of $f$- and $\omega$- Reggeons. Thus, we assume 
$M = M_P + M_f - M_{\omega}$ for $pp$ collisions and $M = M_P + M_f + M_{\omega}$ for $p\bar{p}$ 
 ones. The values of parameters found from a fit to data can be found in~\cite{diff}. The fit result 
is compared with data in Fig.~\ref{Fig:TotalElastic}.\\
The processes of single- and double- diffraction dissociation are closely related to small 
angle elastic scattering in which each of the incoming hadrons may become a system, which 
will then decay into a number of stable final state particles. In high energy physics, 
diffraction is usually defined as any process involving Pomeron exchange. Experimentally, 
there is no possibility to distinguish processes that are caused by Pomeron exchange 
from those that are caused by Reggeon (e.g. $f$-trajectory) exchange. Therefore, we define 
diffraction dissociation as the exchange of both Pomeron and Reggeon. In \cite{diff} it is 
proposed to describe data on soft diffraction dissociation in $pp$ and $p\bar{p}$ 
interactions by taking into account all possible non-enhanced absorptive corrections to 
triple-Regge vertices and loop diagrams (see Fig.~\ref{Fig:SD_DDdiag}). The values of triple-Reggeon coupling 
constants were found from fit to data on double differential single-diffractive cross-section 
versus diffracted mass and transferred momentum. The predictions for integrated single- 
and double- diffractive cross-sections are compared with data in Fig.~\ref{Fig:SD_DD}\\ 
Predictions of the model on total, elastic, single- and double- diffractive cross-sections and on 
the elastic scattering slope ($B=-[d(\ln\sigma_{el} )/dt]_{t=0}$) at various LHC energies are presented 
in Table~\ref{Tab:Tab1}. The uncertainty of the predictions are expected to be ~3$\%$ for total and 
elastic cross-sections and ~10$\%$ for single- and double- diffraction ones.\\
\begin{table}[h!]
\begin{center}
\caption{
\label{Tab:Tab1}
Predictions for LHC.}
\begin{tabular}{cccccc}
\hline
$\sqrt{s}$ TeV & $\sigma_{tot}$ mb & $\sigma_{el}$ mb & $B$ GeV$^{-2}$ & $\sigma_{SD}(M^2 < 0.05s)$ mb &
$\sigma_{DD}(\Delta\eta >3)$ mb \\ 
0.9  &	66.8 &	14.6 &	15.4 &	9.3  &	5.7 \\
2.76 &	81.8 &	19.6 &	17.3 &	11.2 &	5.9 \\
7    &	96.4 &	24.8 &	19   &	12.9 &	6.1 \\
10   &	102  &	27   &	19.8 &	13.6 &	6.2 \\
14   &	108  &	29.5 &	20.5 &	14.3 &	6.4 \\ 
\end{tabular}
\end{center}
\end{table} 
\section*{Acknowledgments}
M.G.P. thanks Jean-Pierre Revol for useful discussions and for suggesting to write this note.

\end{document}